\documentclass[sigconf, screen]{acmart}

\AtBeginDocument{%
  }

\setcopyright{acmlicensed}

\copyrightyear{2026}
\acmYear{2026}
\acmDOI{XXXXXXX.XXXXXXX}
\acmConference[FSE Companion ’26]{FSE Companion ’26, July 5–9, 2026, Montreal, Canada}{July 5–-9, 2026}{Montreal, Canada}
\acmISBN{979-X-XXXX-XXXX-X/26/03}

\acmBooktitle{Companion Proceedings of the 34th ACM Symposium on the Foundations of Software Engineering (FSE '26), June 5--9, 2026, Montreal, Canada}

\usepackage{multirow}

\usepackage{caption}
\usepackage{subcaption}
\usepackage{makecell}
\usepackage{colortbl}
\usepackage{graphicx} 
\usepackage{hhline}
\usepackage{enumitem}
\usepackage{arydshln} 
\usepackage{amsmath}
\usepackage{todonotes}
\usepackage{cleveref}
\usepackage[most]{tcolorbox}
\usepackage{array}
\usepackage{makecell}
\usepackage{fontawesome5}
\usepackage[ruled, vlined]{algorithm2e}
\usepackage[table, dvipsnames]{xcolor}
\usepackage{adjustbox}
\usepackage{capt-of} 
\usepackage{tcolorbox}
\usepackage{tikz}

\usepackage{pifont}

\usetikzlibrary{fit, backgrounds, arrows.meta, positioning, calc}

\tcbuselibrary{breakable,skins}

\newtcolorbox{rqanswer}{
  breakable,
  enhanced,
  colback=gray!5!white,
  colframe=gray!70!black,
  boxrule=0.4pt,
  arc=2pt,
  left=6pt,
  right=6pt,
  top=2pt,
  bottom=2pt,
  fontupper=\small,
}

\newcolumntype{C}[1]{>{\centering\arraybackslash}p{#1}}

\newsavebox{\tablebox}

\usepackage[acronym,nolist,nogroupskip]{glossaries}

\setcopyright{none}
\renewcommand\footnotetextcopyrightpermission[1]{}

\glsdisablehyper

\newcommand{\cmark}{\ding{51}}%
\newcommand{\xmark}{\ding{55}}%

\DeclareMathOperator*{\argmax}{arg\,max}

\definecolor{GreyBlue}{HTML}{8B9DC3}
\definecolor{GreyRed}{HTML}{941F1F}

\newacronym[plural=CFTs, longplural=calls for tenders]{cft}
  {CFT}
  {call for tenders}

\newacronym{nlp}{NLP}{natural language processing}
\newacronym[plural={LLMs}, firstplural={large language models (LLMs)}]{llm}{LLM}{large language model}
\newacronym{cpv}{CPV}{Common Procurement Vocabulary}
\newacronym[plural={GGSs}, firstplural={groups of goods and services (GGSs)}]{ggs}{GGS}{group of goods and services}
\newacronym{foen}{FOEN}{Federal Office for the Environment}
\newacronym{gpp}{GPP}{Green Public Procurement}
\newacronym[plural=SPCs, longplural=sustainable procurement criteria]{spc}{SPC}{sustainable procurement criterion}
\newacronym{gpa}{GPA}{Government Procurement Agreement}
\newacronym{wto}{WTO}{World Trade Organization}

\newacronym[plural=GNs, longplural=generated or LLM-based sustainable procurement criteria]{gn}{GN}{generated or LLM-based sustainable procurement criterion}
\newacronym[plural=GTs, longplural=ground-truth sustainable procurement criteria]{gt}{GT}{ground-truth sustainable procurement criterion}
\newacronym[plural=AAs, longplural=areas of action]{aa}{AA}{Area of Action}
\newacronym[plural=PDs, longplural=Products]{pd}{PD}{Product}
\newacronym[plural=NTs, longplural=Nutrition]{nt}{NT}{Nutrition}
\newacronym[plural=FNs, longplural=Furniture]{fn}{FN}{Furniture}
\newacronym[plural=LMs, longplural=Illumination]{lm}{LM}{Illumination}
\newacronym[plural=TSs, longplural=Transportation]{ts}{TS}{Transportation}
\newacronym[plural=CSs, longplural=Construction]{cs}{CS}{Construction}
\newacronym[plural=PSs, longplural=Printing Services]{ps}{PS}{Printing Service}
\newacronym[plural=CDs, longplural=Communication Devices]{cd}{CD}{Communication Device}
\newacronym[plural={RAGs}, firstplural={retrieval-augmented generations (RAGs)}]{rag}{RAG}{retrieval-augmented generation}

\begin{document}

\title{Generating and Evaluating Sustainable Procurement Criteria for the Swiss Public Sector using In-Context Prompting with Large Language Models}
\renewcommand{\shorttitle}{Generating and Evaluating Procurement Criteria for Swiss Public Procurement using In-context Prompting with LLMs}

\author{Yingqiang Gao}
\authornote{Corresponding authors.}
\orcid{https://orcid.org/0009-0000-0876-621X}
\affiliation{%
  \institution{University of Zurich}
  \city{Zurich}
  \country{Switzerland}
}
\email{yingqiang.gao@uzh.ch}

\author{Veton Matoshi}
\orcid{https://orcid.org/0009-0002-6613-5701}
\affiliation{%
  \institution{Bern University of Applied Sciences}
  \city{Bern}
  \country{Switzerland}
}
\email{veton.matoshi@bfh.ch}

\author{Luca Rolshoven}
\orcid{https://orcid.org/0009-0001-0663-9011}
\affiliation{%
  \institution{Bern University of Applied Sciences}
  \institution{University of Bern}
  \city{Bern}
  \country{Switzerland}}
\email{luca.rolshoven@bfh.ch}

\author{Tilia Ellendorff}
\orcid{https://orcid.org/0000-0002-8543-4902}
\affiliation{%
  \institution{University of Zurich}
  \city{Zurich}
  \country{Switzerland}}
\email{tilia.ellendorff@uzh.ch}

\author{Judith Binder}
\orcid{}
\affiliation{%
  \institution{Bern University of Applied Sciences}
  \city{Bern}
  \country{Switzerland}}
\email{judith.binder@bfh.ch}

\author{Jeremy Austin Jann}
\orcid{https://orcid.org/0009-0001-9840-6142}
\affiliation{%
  \institution{Bern University of Applied Sciences}
  \city{Bern}
  \country{Switzerland}}
\email{jeremy.jann@bfh.ch}

\author{Gerold Schneider}
\orcid{https://orcid.org/0000-0002-1905-6237}
\affiliation{%
  \institution{University of Zurich}
  \city{Zurich}
  \country{Switzerland}}
\email{gerold.schneider@uzh.ch}

\author{Matthias Stürmer}
\authornotemark[1]
\orcid{https://orcid.org/0000-0001-9038-4041}
\affiliation{%
  \institution{Bern University of Applied Sciences}
  \institution{University of Bern}
  \city{Bern}
  \country{Switzerland}}
  \email{matthias.stuermer@bfh.ch}

\renewcommand{\shortauthors}{Y. Gao et al. 2026}

\begin{teaserfigure}
\centering
\hspace{-0.5em}
\includegraphics[width=\textwidth]{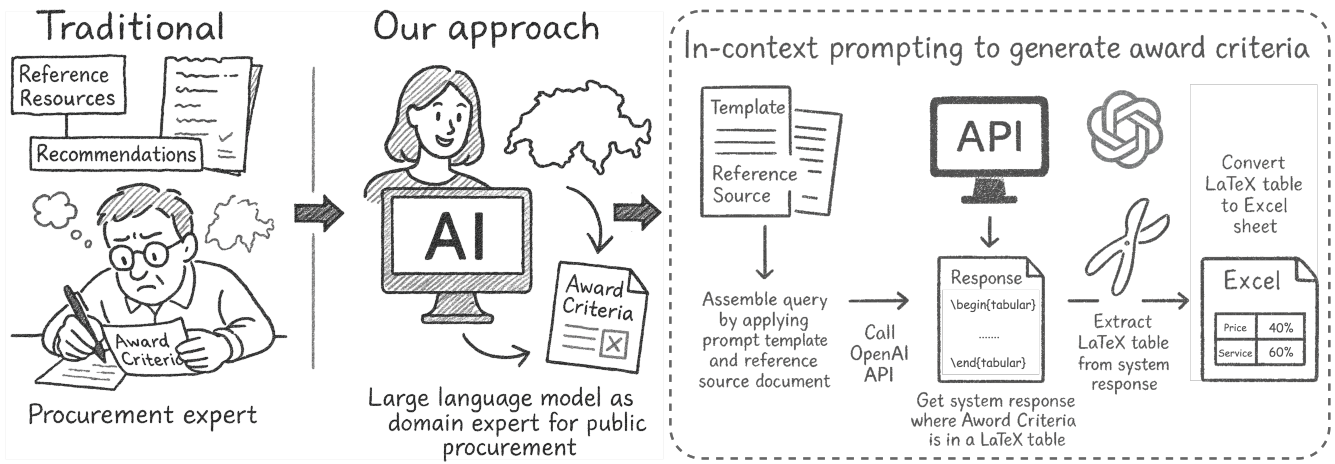}
\caption{Writing sustainable procurement criteria for the Swiss public sector is traditionally a labor-intensive process, requiring substantial human effort and specialized expertise. Our approach leverages state-of-the-art large language models as domain experts to automatically generate sustainable procurement criteria in an end-to-end manner, significantly reducing time and cognitive workload. We highlight the generation of award criteria as an example.}
\end{teaserfigure}

\begin{abstract}

Public procurement refers to the process by which public sector institutions, such as governments, municipalities, and publicly funded bodies, acquire goods and services. Swiss law requires the integration of ecological, social, and economic sustainability requirements into tender evaluations in the format of criteria that have to be fulfilled by a bidder. However, translating high-level sustainability regulations into concrete, verifiable, and sector-specific procurement criteria (such as selection criteria, award criteria, and technical specifications) remains a labor-intensive and error-prone manual task, requiring substantial domain expertise in several groups of goods and services and considerable manual effort. This paper presents a configurable, LLM-assisted pipeline that is presented as a software supporting the systematic generation and evaluation of sustainability-oriented procurement criteria catalogs for Switzerland. The system integrates in-context prompting, interchangeable LLM backends, and automated output validation to enable auditable criteria generation across different procurement sectors. As a proof of concept, we instantiate the pipeline using official sustainability guidelines published by the Swiss government and the European Commission, which are ingested as structured reference documents. We evaluate the system through a combination of automated quality checks, including an LLM-based evaluation component, and expert comparison against a manually curated gold standard. Our results demonstrate that the proposed pipeline can substantially reduce manual drafting effort while producing criteria catalogs that are consistent with official guidelines. We further discuss system limitations, failure modes, and design trade-offs observed during deployment, highlighting key considerations for integrating generative AI into public sector software workflows.

\end{abstract}

\ccsdesc[500]{Applied computing~Computing in government}
\ccsdesc[300]{Social and professional topics~Sustainability}
\ccsdesc[100]{Social and professional topics~Automation}
\ccsdesc[100]{Computing methodologies~Natural language processing}

\keywords{public procurement, sustainability, process automation, large language models, in-context prompting, applied NLP, government decision support, AI governance.}

\maketitle

\pagestyle{plain}

\section{Introduction}

Public procurement represents one of the largest and most regulated decision processes in the public sector, with governments worldwide spending approximately 10 trillion US dollars annually \cite{fazekas_global_2024}. Modern procurement workflows are executed through digital tendering platforms, where requirements, award criteria, and evaluation procedures are specified as structured artifacts that directly shape downstream decision-making \cite{mavidis2025unveiling}. While public tendering is legally mandated to ensure transparency and competition under frameworks such as the \gls{gpa} of the \gls{wto}, the practical implementation of these requirements remains largely manual, relying on procurement officers to translate complex regulatory texts into operationalizable criteria with limited software automation \cite{cao2022chinese, monteiro2023decentralised}.

Public procurement in Switzerland follows legal frameworks to ensure fairness and competition. A typical \gls*{cft} is structured around four requirement types: \textbf{participation conditions} (minimum eligibility, e.g., legal/financial standing), \textbf{selection criteria} (bidder capability, e.g., experience), \textbf{technical specification} (deliverable requirements, e.g., materials, performance), and \textbf{award criteria} (evaluative factors, e.g., price, quality). Traditionally, sustainability had a limited binding impact unless formally embedded in these criteria.

Recent procurement regulations extend beyond purely economic considerations and explicitly require the integration of ecological and social sustainability criteria into \glspl{cft}. The revised \gls{gpa} of 2012 promotes the inclusion of criteria that support environmental protection and resource conservation \cite{koch_green_2020}, formalized in the concept of the ``most economically advantageous tender'' (MEAT; \cite{stilger2017comparative}). In practice, MEAT shifts procurement from a single-objective optimization problem to a multi-criteria decision process, where life-cycle costs, qualitative factors, and sustainability indicators must be simultaneously specified, weighted, and evaluated. However, existing procurement software systems provide limited support for systematically generating and validating such complex criteria sets, leaving these tasks to manual drafting and expert judgment.

\begin{figure}[!t]
    \centering
    \includegraphics[width=\columnwidth]{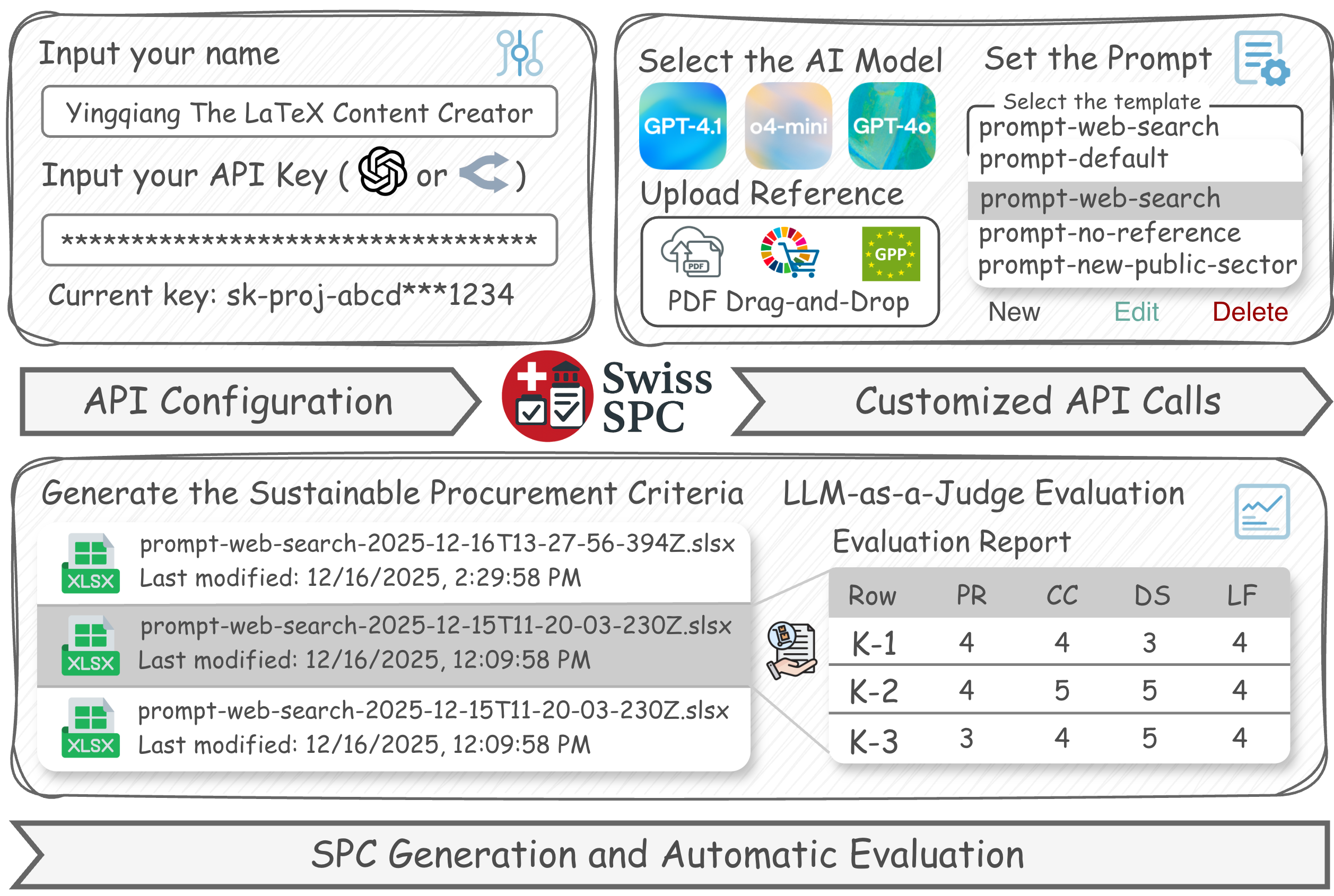}
    \caption{End-to-end pipeline of SwissSPC for sustainable procurement criteria generation and evaluation. SwissSPC integrates API configuration and model selection, prompt templating with optional document references, and XLSX-based criteria generation, followed by LLM-as-a-Judge evaluation to assess generation quality along multiple dimensions.}
    \label{fig:pipeline}
\end{figure}

This gap introduces a recurring challenge. Procurement officers must repeatedly design \glspl{spc} (German \textit{nachhaltige Beschaffungskriterien}) that are legally compliant, sector-specific, and technically verifiable, while balancing cost, quality, and long-term impact. Besides involving a considerable amount of manual work, the high degree of professional discretion involved leads to inconsistent formulations across tenders and organizations, increasing the risk of ambiguity, limited comparability of bids, and legal disputes. Despite the centrality of these criteria to digital procurement workflows, little work has been done to automate their formulation as a software engineering problem \cite{boechat2023public}. 

Building on the identified gap, this paper presents \textbf{SwissSPC}, a desktop software system that operationalizes \glspl{llm} as configurable components within the public procurement workflow to support the automated generation of \glspl{spc}. Rather than treating \glspl{llm} as standalone generators, SwissSPC integrates in-context prompting, document-grounded reference ingestion, and structured output handling to align generated criteria with official regulatory frameworks. This design explicitly addresses the scarcity and fragmentation of domain reference data by anchoring model outputs to authoritative sources published by Swiss and European public institutions.

From a software engineering perspective, SwissSPC targets the automation of a critical yet under-explored task in public procurement: the repeated, manual drafting of sector-specific sustainability criteria that are legally compliant, transparent, and verifiable. By reproducing the structure and content of human-curated criteria catalogs within a controlled software environment, the system reduces cognitive load for procurement officers who are often specialized in particular sectors, and enables faster iteration without sacrificing regulatory alignment. To our knowledge, this work represents the first end-to-end software-based approach that embeds \gls{llm}-driven criteria generation and evaluation into a deployable procurement support tool for Switzerland.

\begin{table*}[htb]
    \centering
    \caption{Structure of an LLM-generated SPC table (other rows are dropped for compact demonstration). Abbreviations: AA (Area of Action), C-ID(Criterion-ID), CC (Criterion Category), SPC (Sustainable Procurement Criterion), AL (Ambition Level).}

    \resizebox{\textwidth}{!}{
        {\renewcommand\arraystretch{1.3} \fontsize{11pt}{13pt}\selectfont
        \begin{tabular}{lcllp{6cm}p{4cm}p{4cm}p{4cm}}
            \toprule
            \textbf{AA} & \textbf{AA-ID} & \textbf{C-ID} & \textbf{CC} & \textbf{SPC} & \textbf{AL: Basis} & \textbf{AL: Good Practice} & \textbf{AL: Exemplary} \\
            \midrule
            \multirow{5}{*}{\makecell[l]{Chemical \\ management}} & \multirow{5}{*}{AA-1} & \multirow{5}{*}{C-01} & \multirow{5}{*}{TS} & The supplier must implement a chemical management system that minimizes the use of hazardous substances. Evidence must be provided through system descriptions and inspection reports. & Proof of a chemical management system. & Proof of a chemical management system with annual review and update. & Proof of a chemical management system with complete elimination of substances of very high concern (SVHC). \\
            \midrule
            \multirow{5}{*}{\makecell[l]{Wood origin \\ and certification}} & \multirow{5}{*}{AA-2} & \multirow{5}{*}{C-02} & \multirow{5}{*}{TS} & The wood used must come from verifiably sustainable forestry. Certificates such as FSC or PEFC must be provided as proof. The traceability of the wood supply chain must be demonstrated through appropriate documentation. & At least 70\% of the wood used is FSC or PEFC certified. & At least 90\% of the wood used is FSC or PEFC certified; complete supply chain documentation. & 100\% of the wood used is FSC or PEFC certified; additionally, proof of chain of custody for all suppliers.\\
            \bottomrule
        \end{tabular}
        }}
    \label{tab:mock-table}
\end{table*}

Beyond system construction, we empirically investigate the feasibility and limitations of this approach through a combination of automated and expert-based evaluation mechanisms. In particular, we examine how LLM-as-a-Judge \citep{zheng2023judging} can be integrated as a scalable quality-assurance component alongside domain expert assessment, and we analyze failure modes that arise when deploying generative models in a high-stakes, regulation-driven software context. This leads to the following research questions (RQs):

\begin{itemize}[noitemsep, partopsep=0pt, parsep=0pt, left=0pt]
\item \textbf{RQ1}: To what extent can an \gls{llm}-based software pipeline reliably generate detailed and sector-specific sustainability procurement criteria from end-to-end?
\item \textbf{RQ2}: How can automated evaluation mechanisms, including LLM-as-a-Judge, be combined with expert assessment to support scalable quality assurance of generated award criteria?
\item \textbf{RQ3}: What engineering challenges and system-level limitations arise when integrating generative AI into real-world public procurement software workflows?
\end{itemize}

Our work consists of three primary contributions. First, we design and implement SwissSPC, an end-to-end, configurable software system that integrates \glspl{llm} into public procurement workflows to automate the generation and evaluation of sustainability-oriented procurement criteria. The generated catalogs are an important resource that serves as a reference to suggest applicable criteria to include when drafting new \glspl{cft} and to be used in downstream applications within the procurement domain. Second, we demonstrate how document-grounded prompting and automated quality-assurance mechanisms, including \gls{llm}-based evaluation, can be leveraged to support auditable, i.e., verifiable by humans or by an LLM-as-a-Judge \citep{llmJudge2023}, criteria formulation in public decision-making processes. Third, through empirical evaluation and system analysis, we identify key engineering challenges, failure modes, and design trade-offs that arise when deploying generative AI in real-world public-sector procurement software, providing actionable insights for both researchers and practitioners.

\section{Related Work}
A range of procurement-related studies has applied \gls{nlp} to \glspl{cft}, including query expansion with semantic linking \citep{Alvarez2011}, BERT-based \cite{devlin2019bert} classification of quality demands for consensus building \citep{Locatelli2023}, and machine learning detection of suspicious tenders \citep{Rabuzin2019,Rabuzin2020}. Moreover, \gls{nlp} techniques have been widely adopted to detect sustainability hints in public procurement and related domains, enabling automated analysis of tenders, contracts, and corporate disclosures. Early efforts relied on keyword matching and tools like Elasticsearch \citep{elasticsearch}
to identify environmental, social, and economic criteria in Belgian procurement documents \cite{Grandia2020}, or used manual and rule-based approaches for Swiss tenders \cite{welz2020sustainability}. However, such keyword-driven methods have proven insufficiently reliable for robust automation \cite{welz2021monitoring}.

Subsequent work adopted more sophisticated \gls{nlp} methods, including supervised classification and embedding-based retrieval, improving the accuracy of identifying sustainability-related content \cite{Torres-Berru2023}. For instance, \citet{Chen2021} leveraged Word2Vec \cite{mikolov2013distributed} and Doc2Vec \cite{le2014distributed} embeddings with SVMs and neural networks to map corporate social responsibility (CSR) reports to UN Sustainable Development Goals (SDGs), achieving over 80\% accuracy. \citet{Endtner2019} use a classification approach with Random Forests to extract selection criteria from Swiss \glspl{cft}, demonstrating automation potential for sustainability integration. More recently, \citet{Matoshi2024} employed \gls{llm}-prompting to extract award criteria from Swiss \glspl{cft}, followed by keyword-based sustainability assessment.

Yet, all these approaches assume the existence of a structured, machine-readable set of \glspl{spc} or keywords (for public procurement), a prerequisite that typically needs human intervention. In practice, \glspl{spc} are then manually derived from official sources such as green procurement handbooks, ecolabel standards, and international frameworks \cite{Andhov2025}. These resources are curated by experts through labor-intensive review of policy documents across sectors like construction, IT, and food services, which is a process that is slow, inconsistent, and difficult to scale.

Crucially, while \glspl{llm} have been used to analyze procurement data, they have not yet been applied to automate the upstream task of extracting and structuring official recommendations for \glspl{spc} themselves. We address this research gap by focusing specifically on the automated extraction of \glspl{spc} from unstructured or semi-structured public guidelines, which is a foundational step that enables downstream analysis but has, to date, remained outside the scope of AI-assisted sustainability monitoring.

\section{Methodology}

\subsection{Data Collection}
\label{sec:data_collection}

To produce the gold standard, a human sustainability expert compiled \glspl{spc} catalogs by collecting source documents from national and international guidelines on sustainable public procurement. Relevant \glspl{spc} were extracted and organized into standardized Excel catalogs, with one catalog per pre-defined \gls{ggs}. One \gls{ggs} covers a sector or partial sector (such as food, furniture, or printing services) with \glspl{ggs} defined in a way that respects thematic coherence within a \gls{ggs}. The identification and delimitation of the different \glspl{ggs} was also performed by the human expert; however, automating this step of the manual workflow is outside of the scope of this paper. For the collection of criteria to be included into a catalog, two main sources were used: (1) the European Union Green Public Procurement (\textcolor{GreyBlue}{EU-GPP})\footnote{\url{https://green-forum.ec.europa.eu/green-business/green-public-procurement_en}}\cite{EU_GPP_General_EC, EU_GPP_Criteria_EC}, which promotes environmental criteria across product life cycles; and (2) the Swiss \textcolor{GreyRed}{Toolbox} for sustainable procurement (\textit{Toolbox Nachhaltige Beschaffung Schweiz} in German)\footnote{\url{https://woeb.swiss/de/toolbox}} \cite{BAFU2025EcologicalProcurement, BAFU2025SustainableProcurementToolbox}, a practice-oriented resource for Swiss public authorities. Each Excel file lists \glspl{spc} with identifiers, categories, \glspl{aa}, sustainability dimensions, and source references, ensuring consistency, traceability, and compatibility with automated processing. Table~\ref{tab:sectors} gives an overview of collected official guidelines used as reference documents when generating the procurement criteria tables. So far we have compiled a gold standard for the following \glspl{ggs}: \gls{nt}~\cite{BAFU2020food,EC2023food, BAFU2020relevanz}, \gls{ts} (excluding railway)~\cite{BAFU2023buses,BAFU2021cars,EC2018road,EC2023lighting,BAFU2020relevanz},  \gls{fn}~\cite{BAFU2023furniture,EC2018furniture, BAFU2020relevanz}, \gls{lm}~\cite{BKB2024circular, EC2023lighting, BAFU2022lighting_indoor, naBe2021lamps}, \gls{ps}~\cite{BAFU2023print,BKB2024circular,EC2020imaging}, \gls{cd}~\cite{BK2024P025ICT,BAFU2023ICTequipment,BKB2024circular,EC2023datacentres}, and \gls{cs}~\cite{KBOB2021designbuild,NNBS2024SNBSbuilding_construction,NNBS2020SNBSinfrastructure,NNBS2024SNBSareal,SBVToolboxConstruction}.

\begin{figure*}[htb]
    \centering
    \includegraphics[width=0.85\linewidth]{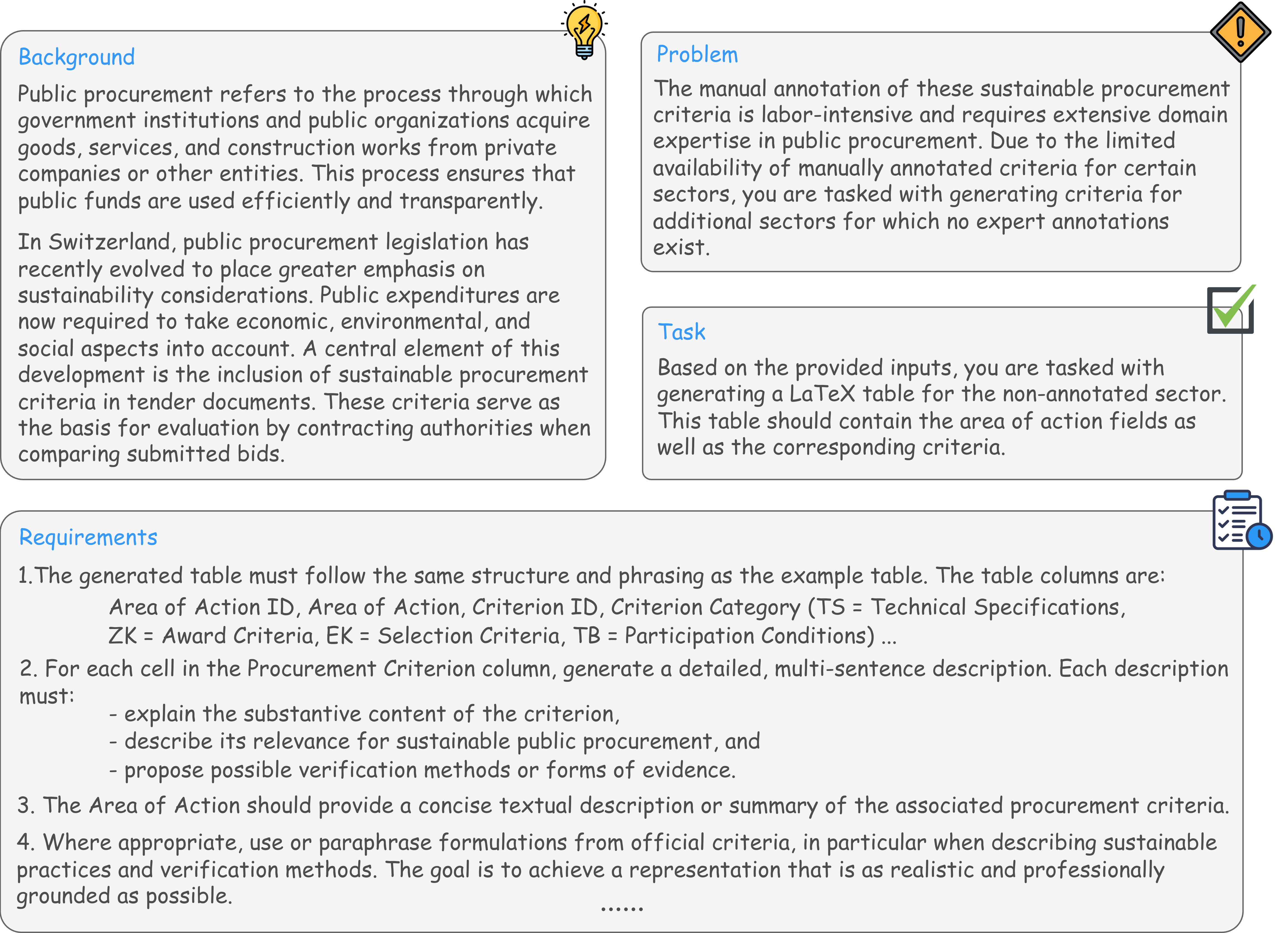}
    \caption{A snippet of the structured prompt used in SwissSPC (translated from German to English) for generating \glspl{spc}. The prompt is decomposed into semantically distinct components: Background, Problem, Task, Requirement, Inputs, etc. (the full prompt contains 875 words), to explicitly encode sector context, motivation, generation objectives, and output constraints.
    }
    \label{fig:prompt-final}
\end{figure*}

\begin{table}[!htb]
    \centering
    \caption{Collected sources for sustainable procurement criteria which serve as references to the \gls{llm} in-context learning.}
    \label{tab:sectors}
    \resizebox{0.9\columnwidth}{!}{
    \begin{tabular}{cccccc}
    \toprule
    \textbf{Sector} & \textbf{EU-GPP} & \textbf{Toolbox} & \textbf{Expert} & \textbf{\# Area} & \textbf{\# Criteria} \\
    \midrule 
    NT & \cmark & \xmark & \cmark & 15 & 79 \\
    FN & \cmark & \cmark & \cmark & 26 & 60 \\
    LM & \cmark & \cmark & \cmark & 52 & 101 \\
    TS &\cmark  & \cmark & \cmark & 69 & 106 \\
    CS & \cmark & \cmark & \cmark & 48 & 144 \\
    PS & \cmark & \cmark & \cmark & 36 & 60 \\
    CD & \cmark & \cmark & \cmark & 61 & 82  \\
    \bottomrule 
    \end{tabular}
    }
\end{table}

\subsection{In-Context Learning}

We formulate the task of generating \glspl{spc} for public procurement as a specialized instance of knowledge acquisition. Our core approach is to apply in-context learning with \glspl{llm} \citep{brown2020language}, treating \glspl{llm} as general-purpose experts capable of transferring sectoral knowledge. Concretely, given a small set of example ground-truth \glspl{spc} (alongside the other information as stated in \ref{sec:data_collection}) $Y$ for a \gls{ggs} $s$ together with the prompt template $C$ including additional context and official reference documents $D_s$, we employ a prompting function $f$ that conditions the \gls{llm} on this information. The model is then prompted to generate \glspl{spc} $\hat{Y}_{s}$, with outputs assessed via a utility function $U(\cdot)$ based on either automatic metrics (e.g., LLM-as-a-Judge \citep{llmJudge2023} or traditional metics based on lexical overlaps) or human evaluation:
\begin{align*}
\hat{Y}_s 
  &= \argmax_{\hat{Y}_s} \, U\!\left[f(D_s, Y_s, C)\right], \text{ where } \\
\hat{y}_{s}^i &\in \hat{Y}_s \sim \mathrm{LLM}\bigl[f(D_s, Y_s, C)\bigr], i=1,\ldots, |\hat{Y}_s|, \hat{y}_s^i \text{ being the token}, \\
U(\cdot) &=
  \begin{cases}
    \mathrm{AutoEval}(\hat{Y}_s), & \text{traditional metrics or LLM-as-a-Judge}, \\
    \mathrm{Manual}(\hat{Y}_s), & \text{human evaluation}.
  \end{cases}
\end{align*}

In practice, recruiting qualified evaluators for all procurement sectors is infeasible, as sector expertise is narrow, costly, and non-scalable. To enable fully automated evaluation, we research the utility function through two routes: LLM-as-a-Judge for rubric-based assessment and automated reference-based lexical and semantic metrics. Human evaluation is retained only for qualitative analysis, focusing on inspecting typical \gls{llm} mistakes. This approach builds on recent findings that evaluator \glspl{llm} can approximate expert judgments with strong agreement when properly prompted and calibrated \citep{kim2024prometheus}.

To acquire and structure domain knowledge, the prompting function $f$ constructs contextualized inputs for zero-shot inference and ensures consistent output formatting. We prompt the \glspl{llm} to generate \glspl{spc} in \LaTeX{} tables aligned with the human annotation format, later converted to Excel sheets for analysis (see Table~\ref{tab:mock-table} for a concrete example). Specifically, the criteria categories include: \textbf{TS} (Technical Specification), \textbf{ZK} (Award Criterion), \textbf{EK} (Selection Criterion), and \textbf{TB} (Participation Conditions), all abbreviated from corresponding German terminologies.

The function $f$ integrates the target sector $s$, prompt template $C$, and reference document $D_s$ as input to the model. We query OpenAI’s model \texttt{gpt-4o}, \texttt{o4-mini}, and \texttt{gpt-4.1} \citep{achiam2023gpt}, chosen for their built-in file search capability that grounds generation in reference documents (i.e., \textcolor{GreyBlue}{EU-GPP} or \textcolor{GreyRed}{Toolbox}). To counter \texttt{gpt-4o}’s tendency to produce short tables, we performed ten generations and selected the most complete output for evaluation. Figure~\ref{fig:prompt-final} demonstrates the original prompt template we used for end-to-end in-context \glspl{spc} generation.

\section{System Design}

\begin{figure*}
    \includegraphics[width=0.9\textwidth]{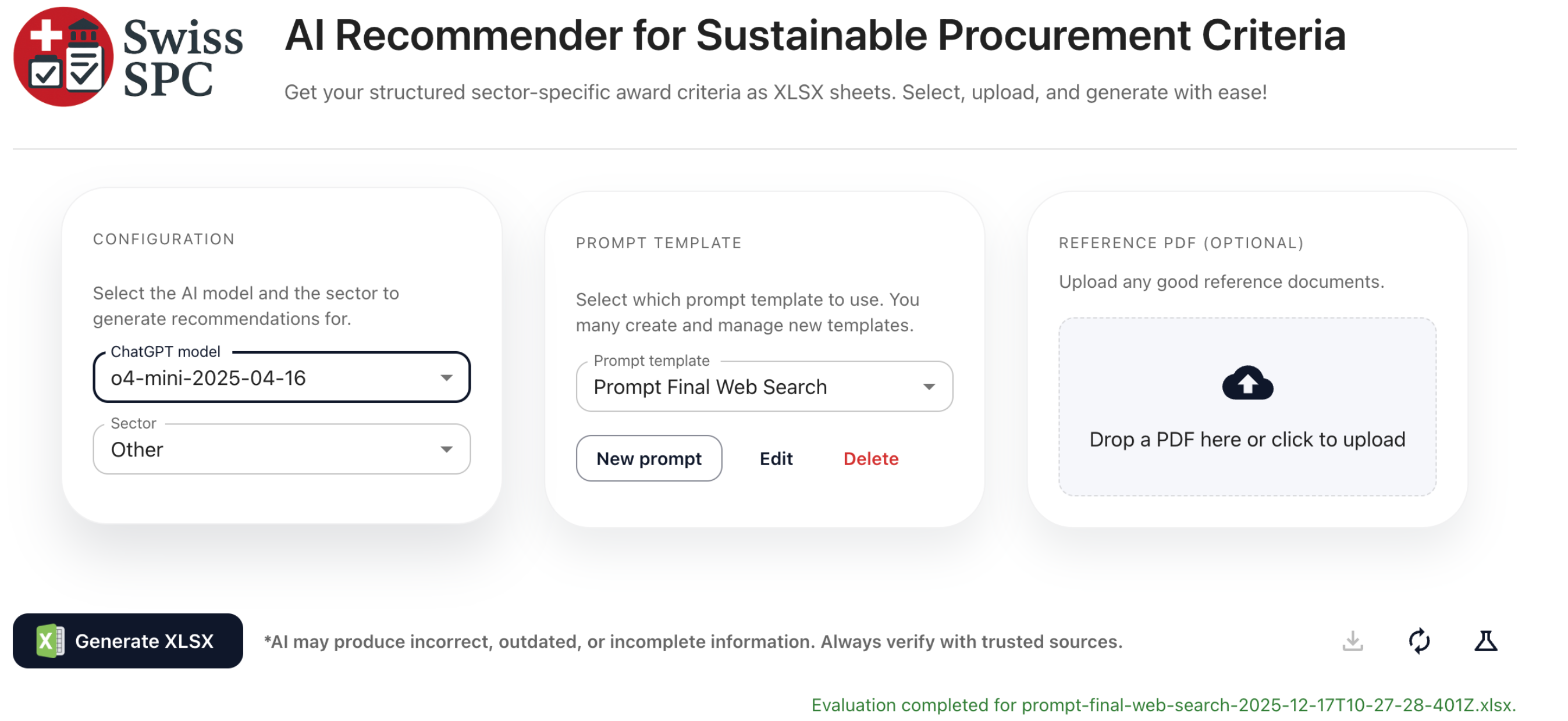}
    \caption{SwissSPC user interface for AI-assisted generation of Sustainable Procurement Criteria. The workflow guides users through (1) generation settings such as model and sector selection, (2) in-context prompting via configurable prompt templates, and (3) optional upload of credited reference PDFs, producing structured, sector-specific \glspl{spc} catalogs exported as XLSX files.}
    \label{fig:screenshot}
\end{figure*}

To address the challenge of scaling the generation of \glspl{spc} across diverse sectors without extensive manual expert annotation, we designed a semi-automated generation pipeline as the workflow of SwissSPC. It follows a \gls{rag} principle \cite{retrievalAugmentedGeneration}, adapted for structured document synthesis. It consists of three primary components: the Context Retrieval Module, the In-Context Prompter, and the XLSX Generator.

\paragraph{\textbf{Context Retrieval Module}}

This module ingests unstructured and semi-structured data sources. Specifically, it processes:
\begin{itemize}[noitemsep, partopsep=0pt, parsep=0pt, left=0pt]
    \item \textbf{Official Recommendations:} PDF parsing of user-uploaded reference documents to extract official sustainability instruments.
    \item \textbf{Expert Recommendations:} Ingestion of existing expert-annotated Excel tables (e.g., the Furniture \gls{ggs}, see Table~\ref{tab:sectors} for more detail) to serve as one-shot examples.
    \item \textbf{Web Search Integration:} A dynamic search component that retrieves external sources (URLs) on the internet to ground generated \glspl{spc}.
\end{itemize}

\paragraph{\textbf{In-Context Prompter}}
The core logic resides in a sophisticated prompt chain designed to enforce strict structural and semantic constraints. The prompt structure includes:
\begin{itemize}[noitemsep, partopsep=0pt, parsep=0pt, left=0pt]
    \item \textbf{Role Definition:} Establishing the persona of an AI procurement expert to align tone and terminology.
    \item \textbf{Structural Constraints:} Enforcing a specific LaTeX tabular schema with defined columns (AA-ID, CC, AL, etc., see the Table~\ref{tab:mock-table} header for more detail).
    \item \textbf{Semantic Requirements:} Mandating multi-sentence, detailed descriptions for \glspl{spc}, ensuring the inclusion of verification methods (e.g., ISO standards, FSC certificates).
    \item \textbf{Ambition Scaling:} Logic to differentiate between \textit{Basis}, \textit{Good Practice}, and \textit{Exemplary} levels, requiring quantitative progression where applicable.
\end{itemize}

\paragraph{\textbf{XLSX Generator}}

The generated output is a raw XLSX file converted from a LaTeX table and containing at least 20 distinct \gls{aa} fields. To support auditability and human-in-the-loop verification, the system enforces a strict source-of-truth requirement, whereby each generated criterion must be accompanied by a valid URL pointing to a trusted public authority or standards body. By fixing the output schema, enforcing ambition-level differentiation, and requiring verifiable external sources, the design limits stylistic variance while preserving sectoral adaptability. To improve robustness under transient API failures or incomplete model responses, the generation process is implemented as a bounded multi-round procedure: if a generation attempt fails to meet structural or source-of-truth constraints, the system automatically retries the request, up to a maximum of five attempts, before reporting an error.

The system currently relies on high-capacity closed-source large language models provided by OpenAI to generate the final criteria artifacts. This design choice is motivated by engineering constraints observed in early prototyping, including the need for long-context handling, stable instruction following, and consistent structured output generation when grounding responses in regulatory reference documents. 

At the time of system development, open-source alternatives exhibited higher variance in output structure and lower reliability under document-grounded prompting, which posed challenges for transparency and downstream validation in a safety- and compliance-critical domain such as public procurement.

\begin{algorithm}[!htb]
\caption{Bipartite Matching Table Evaluation}
\label{alg:bipartite-evaluation}

\KwIn{Generated table $T_{GN}$, ground-truth table $T_{GT}$}
\KwOut{Quality, coverage, harmonic score (per-column)}

\SetKwProg{Fn}{Function}{:}{}
\SetKwFunction{BERTScore}{BERTScore}
\SetKwFunction{Hungarian}{Hungarian}

\BlankLine

\tcp{Stage 1: Build Similarity Matrix for Matching}
Create matrix $M$ of size $|T_{GN}| \times |T_{GT}|$\;
\For{$i = 1$ \KwTo $|T_{GN}|$}{
    \For{$j = 1$ \KwTo $|T_{GT}|$}{
        $M[i,j] \leftarrow$ \BERTScore{$T_{GN}[i][\text{AA}]$, $T_{GT}[j][\text{AA}]$}\;
    }
}

\BlankLine

\tcp{Stage 2: Find Optimal Assignment}
$\text{pairs} \leftarrow$ \Hungarian{$M$} \tcp{Returns optimal (hyp, ref) pairs}

\BlankLine

\tcp{Stage 3: Compute Scores for Matched Pairs Only}
\tcp{AA scores: reuse from matching matrix}
$S_{\text{AA}} \leftarrow [M[i,j] : (i,j) \in \text{pairs}]$\;

\tcp{SPC scores: compute for matched pairs}
$S_{\text{SPC}} \leftarrow []$\;
\For{each pair $(i,j) \in \text{pairs}$}{
    $s \leftarrow$ \BERTScore{$T_{GN}[i][\text{SPC}]$, $T_{GT}[j][\text{SPC}]$}\;
    Append $s$ to $S_{\text{SPC}}$\;
}

\BlankLine

\tcp{Stage 4: Calculate Metrics}
$\text{coverage} \leftarrow \frac{|T_{GN}|}{|T_{GT}|}$ \tcp{Fraction of ground truth covered}

\tcp{Column-specific metrics}
\For{$c \in \{\text{AA}, \text{SPC}\}$}{
    $\text{quality}_c \leftarrow \text{mean}(S_c)$ \tcp{Average BERTScore for column}
    $\text{harmonic}_c \leftarrow \frac{2 \times \text{quality}_c \times \text{coverage}}{\text{quality}_c + \text{coverage}}$\;
}

\KwRet{Column-specific: quality, coverage, harmonic scores}

\end{algorithm}

\section{Quality Assessment}
\subsection{Qualitative Evaluation}

\subsubsection{\textbf{Evaluation Setup}}

The qualitative evaluation was conducted by two data scientists familiar with Swiss public procurement, comparing the generated and ground-truth \glspl{spc} catalogs for the \gls{fn} \gls{ggs}. They focused on the catalog generated from the \textcolor{GreyRed}{Toolbox} source, prioritizing national over international recommendations and assessing only the alignment of \glspl{spc}, which were deemed the most relevant catalog component. With 50 \glspl{gn} and 60 \glspl{gt}, 3,000 criterion pairs were compared. To make this feasible, cosine similarities between each \gls{gn} and all \glspl{gt} using the multilingual mE5\textsubscript{Large} embeddings \citep{wang2024multilingual} were computed and ranked. This enabled a focused evaluation of each \gls{gn} individually, where similar pairs appeared early, allowing evaluators to skip unrelated \glspl{gt} after repeated non-matches. Additional filtering by thematic \gls{aa} further reduced comparisons. The following labels were used to differentiate different matching cases: \textbf{GN = GT} (full equivalence), \textbf{GN $\subset$ GT} (GN covered by GT), \textbf{GT $\subset$ GN} (GT covered by GN), \textbf{GN $\cap$ GT} (partial overlap), and \textbf{GN $\neq$ GT} (no overlap, including skipped cases).

\begin{table*}[!htb]
\centering
\caption{Automatic evaluation results. Generated tables are evaluated against reference sources, with results based on \textcolor{GreyBlue}{EU-GPP} recommendations shown in \textcolor{GreyBlue}{blue} and those based on \textcolor{GreyRed}{Toolbox} in \textcolor{GreyRed}{red}. Cases with identical performance across both sources are marked in black. For each instance, only the reference source yielding the better performance is reported.}
\label{tab:auto_eval_compact}
\begin{tabular}{l l C{1.2cm} C{1.2cm} C{1.2cm} c C{1.6cm} C{1.6cm} C{1.6cm}}
\hline
\multirow{2.5}{*}{\textbf{Sector}} & \multirow{2.5}{*}{\textbf{Model}} &
\multicolumn{3}{c}{\textbf{AA (Area of Action)}} & &
\multicolumn{3}{c}{\textbf{SPC (Sustainable Procurement Criteria)}} \\
\cmidrule(lr){3-5}\cmidrule(lr){7-9}
& &
\textbf{Coverage} & \textbf{Quality} & \textbf{Harmonic} & &
\textbf{Coverage} & \textbf{Quality} & \textbf{Harmonic} \\
\hline
\multirow{3}{*}{NT} & \texttt{gpt-4.1} & \textcolor{GreyBlue}{0.6329} & \textcolor{GreyBlue}{0.2159} & \textcolor{GreyBlue}{0.3220} & & \textcolor{GreyBlue}{0.6359} & \textcolor{GreyBlue}{0.6540} & \textcolor{GreyBlue}{0.6433} \\
& \texttt{gpt-4o}  &  \textcolor{GreyBlue}{0.1519} & \textcolor{GreyBlue}{0.7013} & \textcolor{GreyBlue}{0.2497} & & \textcolor{GreyBlue}{0.1519} & \textcolor{GreyBlue}{0.6648}  & \textcolor{GreyBlue}{0.2473}  \\
   & \texttt{o4-mini} & \textcolor{GreyBlue}{0.5063} & \textcolor{GreyBlue}{0.2770} & \textcolor{GreyBlue}{0.3581} & & \textcolor{GreyBlue}{0.5063} & \textcolor{GreyBlue}{0.6465} & \textcolor{GreyBlue}{0.5679}  \\
\hline
\multirow{3}{*}{FN} & \texttt{gpt-4.1} & 0.8333 & \textcolor{GreyBlue}{0.3892} & \textcolor{GreyBlue}{0.5305} & & 0.8333 & \textcolor{GreyBlue}{0.6619} & \textcolor{GreyBlue}{0.7378} \\
   & \texttt{gpt-4o}  & \textcolor{GreyRed}{0.2500}  & \textcolor{GreyRed}{0.8435} & \textcolor{GreyRed}{0.3857} & & \textcolor{GreyRed}{0.2500} & \textcolor{GreyRed}{0.7514}  &  \textcolor{GreyRed}{0.3752} \\
   & \texttt{o4-mini} & 0.3333 & \textcolor{GreyRed}{0.7568} & \textcolor{GreyRed}{0.4628} & & 0.3333 & \textcolor{GreyRed}{0.6703} & \textcolor{GreyRed}{0.4452} \\
\hline
\multirow{3}{*}{LM} & \texttt{gpt-4.1} & 0.4950 & \textcolor{GreyBlue}{0.7067} & \textcolor{GreyBlue}{0.5822} & & 0.4950 & \textcolor{GreyBlue}{0.6533} & \textcolor{GreyBlue}{0.5633} \\
   & \texttt{gpt-4o}  & 0.1386 & \textcolor{GreyBlue}{0.7465} & \textcolor{GreyBlue}{0.2338} & & 0.1386 & \textcolor{GreyBlue}{0.6703} & \textcolor{GreyBlue}{0.2297} \\
   & \texttt{o4-mini} & 0.1980 & \textcolor{GreyRed}{0.7250} & \textcolor{GreyRed}{0.3111} & & 0.1980 & \textcolor{GreyRed}{0.6728} & \textcolor{GreyRed}{0.3060} \\
\hline
\multirow{3}{*}{TS} & \texttt{gpt-4.1} & 0.4717 & \textcolor{GreyBlue}{0.7228} & \textcolor{GreyBlue}{0.5708} & & 0.4717 & \textcolor{GreyRed}{0.6491} & \textcolor{GreyRed}{0.5464} \\
   & \texttt{gpt-4o} & \textcolor{GreyRed}{0.1698} & \textcolor{GreyBlue}{0.7600} & \textcolor{GreyRed}{0.2768} & & \textcolor{GreyRed}{0.1698} & \textcolor{GreyRed}{0.6492} & \textcolor{GreyRed}{0.2692} \\
   & \texttt{o4-mini} & \textcolor{GreyRed}{0.4717} & \textcolor{GreyRed}{0.7306} & \textcolor{GreyRed}{0.5733} & & \textcolor{GreyRed}{0.4717} & \textcolor{GreyRed}{0.6489} & \textcolor{GreyRed}{0.5463} \\
\hline
\multirow{3}{*}{CS} & \texttt{gpt-4.1} & 0.3472 & \textcolor{GreyBlue}{0.7579} & \textcolor{GreyBlue}{0.4763} & & 0.3472 & \textcolor{GreyBlue}{0.6292} & \textcolor{GreyBlue}{0.4475} \\
   & \texttt{gpt-4o} & 0.0694 & \textcolor{GreyBlue}{0.8235} & \textcolor{GreyBlue}{0.1281} & & 0.0694 & \textcolor{GreyBlue}{0.6274} & 0.1250 \\
  & \texttt{o4-mini} & \textcolor{GreyBlue}{0.6944}  & \textcolor{GreyRed}{0.7668} & \textcolor{GreyBlue}{0.4991} & & \textcolor{GreyBlue}{0.6944} & \textcolor{GreyRed}{0.6233} & \textcolor{GreyBlue}{0.6519} \\
\hline
\multirow{3}{*}{PS} & \texttt{gpt-4.1} & 0.8333 & \textcolor{GreyBlue}{0.5344} & \textcolor{GreyBlue}{0.6512} & & 0.8333 & \textcolor{GreyRed}{0.6539} & \textcolor{GreyRed}{0.7328} \\
   & \texttt{gpt-4o}  & \textcolor{GreyRed}{0.2667} & \textcolor{GreyRed}{0.7806} & \textcolor{GreyRed}{0.3937} & & \textcolor{GreyRed}{0.2667} & \textcolor{GreyRed}{0.6653} & \textcolor{GreyRed}{0.3807} \\
   & \texttt{o4-mini} & \textcolor{GreyRed}{0.8333} & \textcolor{GreyBlue}{0.7687} & \textcolor{GreyRed}{0.6491} & & \textcolor{GreyRed}{0.8333} & \textcolor{GreyBlue}{0.6506} & \textcolor{GreyRed}{0.7301} \\
\hline
\multirow{3}{*}{CD} & \texttt{gpt-4.1} & 0.6098 & \textcolor{GreyBlue}{0.6915} & \textcolor{GreyBlue}{0.6481} & & 0.6098 & \textcolor{GreyRed}{0.6465} & \textcolor{GreyRed}{0.6276} \\
   & \texttt{gpt-4o}  & 0.1220 & \textcolor{GreyBlue}{0.7289} & \textcolor{GreyBlue}{0.2089} & & 0.1220 & \textcolor{GreyRed}{0.6356} & \textcolor{GreyRed}{0.2046} \\
   & \texttt{o4-mini} & \textcolor{GreyBlue}{1.0000} & \textcolor{GreyRed}{0.6822} & \textcolor{GreyBlue}{0.6823} & & \textcolor{GreyBlue}{1.0000}  & \textcolor{GreyRed}{0.6461} & \textcolor{GreyBlue}{0.7723} \\
\hline
\end{tabular}%
\end{table*}

\subsubsection{\textbf{Inter-Annotator Agreement}}\label{sec:agreement}

The original agreement between the two annotators was 94.67\%, indicating high consistency in clear-cut cases. However, Cohen’s Kappa \citep{cohen1960coefficient} was 0.520, reflecting only moderate agreement beyond chance. This discrepancy suggests that while the raters generally agreed on obvious matches and non-matches, the distinction between nuanced categories, particularly partial overlaps ($\mathrm{GN} \cap \mathrm{GT}$) and subset relations ($\mathrm{GN} \subset \mathrm{GT}$,\ \text{and}\ $\mathrm{GT} \subset \mathrm{GN}$), introduced subjectivity into the labeling process. After the independent annotation process, the two annotators discussed disagreements to reach consensus and support a detailed qualitative analysis.

\subsubsection{\textbf{Findings}}\label{sec:human-eval-findings}
The \glspl{gn} successfully captured most of the \glspl{gt}, with corresponding generated criteria identified for the vast majority of ground truth entries, including cases of partial and full overlap. Only 8 \glspl{gt} remained without any corresponding \glspl{gn}. The qualitative analysis of the \glspl{gn} revealed several systematic patterns in content, specificity, and structure.
First, the \glspl{gn} are generally more generic than the \glspl{gt}, a difference that is mirrored in their average lengths: 148.58 characters for \glspl{gn} compared to 483.92 characters for \glspl{gt}.
Some \glspl{gn} are very vague; for example, the following \gls{gn} lacks concrete, enforceable requirements: \textit{"The supplier must provide transparent reporting on sustainability aspects. Evidence must be provided through sustainability reports or GRI reports."} (GN-33).

Second, while \glspl{gt} frequently reference specific certifications (e.g., \textit{ISO}, \textit{EU Ecolabel}), the \glspl{gn} often use general formulations, such as \textit{“Evidence via material balances and efficiency metrics”} (GN-35).

Finally, the \glspl{gn} included sustainability dimensions absent from the \glspl{gt}, e.g., anti-corruption and data privacy, suggesting that the model can surface relevant but underrepresented aspects. Nevertheless, these topics were not covered by the Toolbox source used to generate the \glspl{gn}.
A structural pattern emerged: pairs of related \glspl{gn}, where the first defines a requirement and the second specifies its evaluation. For example: GN-41 \textit{``The supplier must demonstrate measures to reduce greenhouse gas emissions.''} and GN-42 \textit{``Evaluation is based on the quality and scope of climate protection measures.''}

\subsection{Automatic Evaluation}

\subsubsection{\textbf{Traditional Metrics.}}

Our evaluation framework uses bipartite matching to measure the quality of \glspl{gn} against \glspl{gt}. It provides complementary metrics for semantic accuracy and content coverage. Besides \glspl{spc}, we also evaluate the generated \glspl{aa}, where we design a holistic approach to solve the alignment problem of \glspl{gn} and \glspl{gt} (see Algorithm~\ref{alg:bipartite-evaluation}).

We first construct a BERTScore-based \citep{zhang2020bertscore} similarity matrix between all generated and ground-truth rows using the \gls{aa} column as the semantic anchor for exhaustive alignment, since AA encodes the core topical information. We set the language of BERTScore to German and this essentially assembles a multilingual RoBERTa \cite{liu2019roberta} model for loading token embeddings. The Hungarian algorithm \cite{kuhn1955hungarian} then identifies the optimal one-to-one alignment that maximizes overall similarity. For each matched pair, we compute BERTScore for both \gls{aa} and \gls{spc} columns where we reuse previously computed scores for efficiency. We define three metrics:
\begin{itemize}[noitemsep, partopsep=0pt, parsep=0pt, left=0pt]
    \item \textbf{Quality}: Averaged BERTScore similarity between matched rows, reflecting semantic alignment;
    \item \textbf{Coverage}: Ratio of generated to reference rows, indicating completeness;
    \item \textbf{Harmonic Score}: Harmonic mean of quality and coverage, balancing accuracy and completeness;
\end{itemize}
To ensure scalability, the framework batches BERTScore computations and limits comparisons to matched pairs only, reducing complexity for \glspl{spc} from $\mathcal{O}(|T_{GN}|\times|T_{GT}|)$ to $\mathcal{O}(|T_{GN}|)$ without compromising evaluation fidelity (see results in Table~\ref{tab:auto_eval_compact}).

\subsubsection{\textbf{LLM-as-a-Judge.}}
We design an LLM-as-a-Judge framework to assess the practical quality and usability of \glspl{gn} from a domain expert perspective, overcoming the limitations of similarity-based metrics that fail to capture the real-world quality of \glspl{spc}.

\begin{figure*}[]
    \centering
    \includegraphics[width=\linewidth]{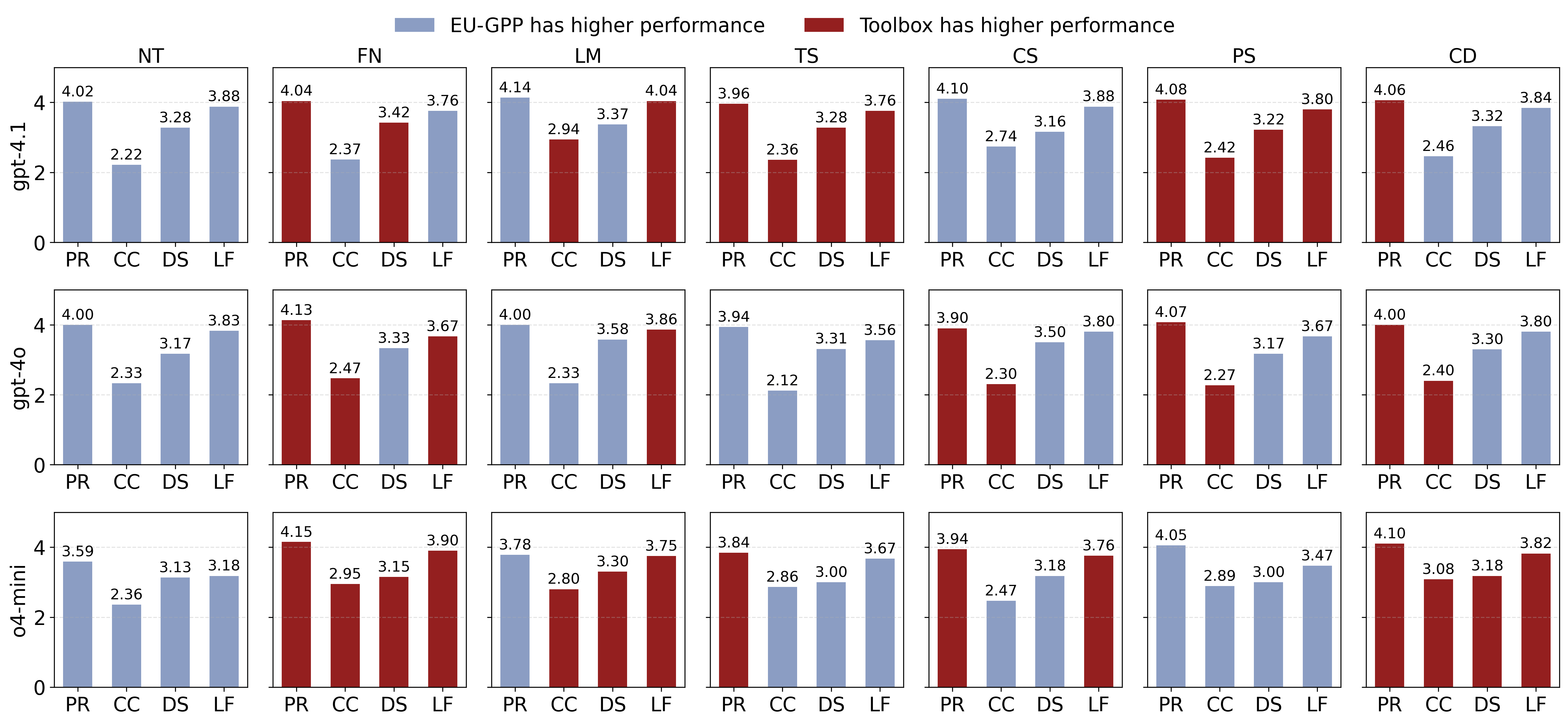}
    \caption{LLM-as-a-Judge evaluation scores for \glspl{spc}. We report the better performing scores using either \textcolor{GreyBlue}{EU-GPP} or \textcolor{GreyRed}{Toolbox} when generating the outputs. Abbreviations: PR (Precision \& Relevance), CC (Completeness \& Coverage), DS (Differentiation \& Scalability), LF (Language \& Formality).}
    \label{fig:LLM-as-a-Judge}
\end{figure*}

Our system uses \texttt{GPT-5-mini} \cite{openai2025gpt5} as a domain-specialized evaluator prompted with the standards and terminology of Swiss sustainable procurement (see Figure~\ref{fig:llm-judge-prompt-en} for more detail). It scores \glspl{gn} along four key dimensions from 1 to 5:
\begin{itemize}[noitemsep, partopsep=0pt, parsep=0pt, left=0pt]
\item \textbf{Precision and Relevance}: Alignment with Swiss best practices and sector relevance;
\item \textbf{Completeness and Coverage}: Clarity on what is required, why it matters, and how it can be demonstrated;
\item \textbf{Differentiation and Scalability}: Presence of clear, measurable distinctions;
\item \textbf{Language and Formality}: Professional tone, correct formulation, and suitable terminology.
\end{itemize}

We have implemented standardized prompts, structured outputs, and automated error handling to ensure consistency and robustness across all evaluations. Figure~\ref{fig:LLM-as-a-Judge} presents the evaluation results obtained with our LLM-as-a-Judge framework, showing average scores across all generated \glspl{spc}.

\begin{tcolorbox}[width=\columnwidth,
  enhanced jigsaw,
  breakable,
  colback=gray!5!white,
  colframe=gray!75!black,
  fontupper=\small,
]

You are an expert in sustainable public procurement in Switzerland and are asked to assess automatically generated sustainable procurement criteria along multiple evaluation dimensions. Please carefully read the following row from a table and evaluate the quality of the formulation of the criterion as well as the associated ambition levels (Basis, Good Practice, Exemplary) according to the aspects listed below:

\medskip

\textbf{Evaluation Dimensions}:
\begin{enumerate}[noitemsep, partopsep=0pt, parsep=0pt, left=0pt]
    \item \textit{Technical correctness and relevance}: Does the content reflect the current state of knowledge on sustainable public procurement in Switzerland? Is the topic relevant for the respective sector?
    \item \textit{Completeness of the description}: Does the criterion answer the questions: What is required? Why is it relevant? How can compliance be demonstrated? Does the criterion consist of at least three technically clear and coherent sentences?
    \item \textit{Ambition levels – differentiation and scalability}: Are the three ambition levels meaningfully differentiated and formulated with progressively increasing requirements? Are there understandable quantitative or qualitative differences between ``Basis'', ``Good Practice'', and ``Exemplary''?
    \item \textit{Linguistic and formal quality}: Is the formulation professional, formally correct, and suitable for official tender documents? Are unnecessary boilerplate phrases avoided and technical terminology used correctly?
\end{enumerate}

\textbf{Your Task}:
Evaluate the entire table row on a scale from 1 (insufficient) to 5 (excellent) for each of the dimensions listed above. In addition, provide a brief justification (1--3 sentences) for each rating. Optionally, suggest improvements to the formulation if any weaknesses are identified.

\textbf{Response Format}:

Area of Action ID: HF-XX  

\medskip
Criterion ID: K-XX
\begin{enumerate}[noitemsep, partopsep=0pt, parsep=0pt, left=0pt]
    \item Technical correctness and relevance: [1--5]  
    Justification: [...]
    \item Completeness of the description: [1--5]  
    Justification: [...]
    \item Ambition levels – differentiation and scalability: [1--5]  
    Justification: [...]
    \item Linguistic and formal quality: [1--5] 
    Justification: [...]
    \item (Optional) Suggested improvement: [...]
\end{enumerate}

\textbf{Please assess the following table row}:
\{ current row  \}

\end{tcolorbox}
\vspace{-1em}
\captionof{figure}{Prompt used in SwissSPC for automated quality assurance of generated sustainable procurement criteria. The prompt instructs the model to evaluate individual criteria along four structured dimensions.}
\vspace{1em}
\label{fig:llm-judge-prompt-en}

\subsubsection{\textbf{Findings}}

Across all evaluated \glspl{ggs}, \texttt{gpt-4.1} consistently demonstrated the most balanced performance between coverage and semantic quality, yielding the highest harmonic scores overall (Table~\ref{tab:auto_eval_compact}). Its outputs closely mirrored the structure and conceptual intent of expert-authored \glspl{spc} catalogs, indicating its ability to internalize and reproduce regulatory language and sustainability logic. In contrast, \texttt{gpt-4o} generated highly precise but less comprehensive results, reflecting a strong adherence to official phrasing yet weaker generalization across diverse procurement domains. \texttt{o4-mini}, while smaller, performed competitively in \glspl{ggs} characterized by well-defined sustainability terminology (e.g., construction and communication devices), underscoring the effectiveness of domain-specific lexicons in guiding constrained models.

A key finding is the influence of source document standardization on model behavior. \glspl{spc} generated from EU-GPP recommendations exhibited higher textual precision and conceptual consistency, attributable to the formalized and repetitive structure of EU guidelines. Conversely, \glspl{spc} derived from the Swiss Toolbox displayed greater topical coverage and contextual variety, suggesting that heterogeneous training input stimulates broader but more diffuse model reasoning. This trade-off between precision and completeness aligns with prior observations in retrieval-based generation, where uniform templates facilitate faithful reproduction but limit creative generalization.

\section{Lessons Learned}

\subsection{Reliability of \gls{llm}-generated \glspl{spc} and LLM-as-a-Judge Evaluations}

The LLM-as-a-Judge evaluation and qualitative human assessment converged on similar patterns. While most generated criteria captured the intent of corresponding expert entries, the \glspl{llm} often produced shorter and more generic formulations, averaging roughly one-third the length of ground-truth criteria. Such conciseness occasionally compromised specificity, particularly when certification standards or quantitative thresholds were expected. Nonetheless, evaluators noted that several generated entries introduced emergent sustainability dimensions, including data privacy, ethical sourcing, and anti-corruption, that were absent from the official catalogs. This indicates that in-context prompting can surface latent sustainability knowledge implicitly encoded in the pretraining data, potentially enriching static expert frameworks.

Structurally, the \glspl{llm} exhibited a notable pattern of pairing requirement-evaluation statements (e.g., defining an obligation followed by an assessment criterion). This emergent consistency suggests that the model not only replicates linguistic surface patterns but also approximates the procedural reasoning behind regulatory text construction. However, observed deviations such as vague evidence descriptions or incomplete ambition-level hierarchies highlight the need for post-generation expert review before policy adoption.

Our results demonstrate that in-context prompting provides a viable pathway toward semi-automated generation of sustainable procurement catalogs. It bridges the gap between costly manual drafting and large-scale reproducibility, while maintaining interpretability and traceability required in governmental decision-support systems. Although current outputs fall short of full expert adequacy, the framework effectively transforms static sustainability guidelines into dynamic, model-assisted resources. This marks a concrete step toward integrating generative AI into automated workflows of public procurement authorities, promoting consistency, transparency, and inclusivity in sustainability-oriented tendering.

\begin{rqanswer}
\textbf{Answer to RQ1:} We show that the proposed \gls{llm}-based pipeline can generate sector-specific sustainability procurement criteria with a high degree of structural consistency and regulatory alignment. While the generated catalogs closely match expert-curated references in terms of content coverage and formulation style, occasional omissions and over-generalizations remain, particularly in highly specialized procurement domains. These findings indicate that automated generation is feasible as a decision-support mechanism, but still benefits from human verification in safety-critical settings.
\end{rqanswer}

\subsection{System Limitations, Design Trade-offs, and Failure Modes}

Despite demonstrating the feasibility of automating sustainability-oriented criteria generation, SwissSPC exhibits several limitations inherent to the use of generative models in a regulation-driven software context. First, the quality of generated criteria remains dependent on the coverage and specificity of the underlying reference documents. In procurement domains where official sustainability guidelines are sparse, fragmented, or highly abstract, the system may produce criteria that are formally correct but lack sufficient operational detail, requiring additional human refinement.

A central design trade-off in SwissSPC concerns robustness versus determinism. To mitigate transient API failures, incomplete responses, and violations of structural constraints, the system adopts a bounded multi-round generation strategy with a fixed retry budget. While this approach substantially improves reliability, it introduces controlled non-determinism in content realization and increases execution latency and API costs. We deliberately prioritize robustness and schema compliance over single-shot execution, reflecting the requirements of safety- and compliance-critical public procurement workflows.

The system also exhibits characteristic failure modes associated with \glspl{llm}. These include the generation of superficially plausible but weakly grounded criteria, over-generalization across \glspl{ggs}, and occasional repetition or redundancy across ambition levels. Although the enforced source-of-truth requirement reduces the risk of hallucinated references, it does not fully eliminate the need for human-in-the-loop verification, particularly when interpreting nuanced regulatory language or sector-specific technical constraints.

Finally, the current implementation relies on high-capacity closed-source \glspl{llm} to meet engineering requirements related to long-context handling and output stability. While the system architecture is model-agnostic by design, substituting open-source models may require additional tuning or post-processing to achieve comparable reliability. These limitations underscore that SwissSPC is best positioned as a decision-support system rather than a fully autonomous procurement tool, and they motivate future work on hybrid human--AI workflows and improved validation mechanisms.

\begin{rqanswer}
    
\textbf{Answer to RQ2}: LLM-as-a-Judge can support scalable quality assurance by pre-screening generated \glspl{spc} for structural validity, consistency, and surface-level grounding, thereby reducing expert workload. However, due to generative failure modes such as over-generalization and weak regulatory grounding, expert assessment remains essential for interpreting nuanced requirements and ensuring enforceability. A staged human-in-the-loop workflow that combines automated screening with targeted expert review enables scalability while preserving the rigor required in compliance-critical procurement settings.
\end{rqanswer}

\subsection{User Study with Procurement Expert}

We conducted a formative user study with a domain expert in sustainable public procurement, who used SwissSPC to generate criteria tables from official sustainability documents and compared the outputs against manually curated \glspl{spc} used in professional practice.

The expert confirmed that SwissSPC captures the overall structural logic of \glspl{spc}: after interacting with the software, most required Excel columns were instantiated correctly and the resulting tables largely conformed to the expected schema. Generated \glspl{aa} and \glspl{spc} were described as internally coherent and plausible, indicating that the system provides a useful structural scaffold, particularly in early stages of criteria development, which can facilitate the traditional workflow

However, several limitations restrict professional usability. The system frequently paraphrases or summarizes criteria instead of reproducing source text verbatim, undermining traceability and legal precision and creating substantial post-editing effort. Generated \glspl{spc} were also incomplete, with fewer \glspl{aa}, missing criteria, and omitted ambition levels compared to manual annotations. In addition, the system occasionally hallucinated numerical thresholds or ambition levels, which is especially problematic in compliance-critical procurement contexts. At the workflow level, limited support for multi-document processing and occasional generation instability further reduced efficiency.

Based on this feedback, key improvement directions include enforcing a strict verbatim extraction mode with explicit source traceability, as well as supporting predefined structure mapping between document sections and \gls{spc} columns. We plan to integrate more effective \gls{rag} designs into the next version of SwissSPC. 

\begin{rqanswer}
\textbf{Answer to RQ3}: The expert evaluation highlights a fundamental mismatch between generative AI and procurement requirements for precision and traceability, as models tend to paraphrase or hallucinate instead of reproducing source text verbatim. Weak source grounding, limited multi-document support, and low user control further reduce robustness, indicating the need for constrained, validation-driven extraction rather than open-ended generation.
\end{rqanswer}

The user study highlights that while SwissSPC shows promising structural understanding of \glspl{spc}, achieving professional-grade usability requires prioritizing precision, completeness, and transparency over generative flexibility.

\section{Outlook and Future Work}

Our central outlook is to develop an aspect-based evaluation framework to assess \glspl{gn} along dimensions such as environmental, social, and economic sustainability, 
thereby enabling a more fine-grained evaluation. We also plan to expand human evaluations with domain experts from sustainability, procurement, and policy to assess the validity and applicability of \glspl{gn} in real procurement contexts. On the technical side, we aim to automate the retrieval of authoritative sustainability sources and enhance the generation process through improved prompting and reasoning strategies, including agentic frameworks that enable models to justify and iteratively refine their outputs. 
Furthermore, the approach will be extended to compile \glspl{spc} catalogs for \glspl{ggs} where official recommendations are unavailable or scarce, thereby supporting sectors that currently lack structured sustainability guidance. 
Finally, we envision expanding our approach toward the discovery of emerging sustainability criteria by leveraging unofficial or pre-standardized sources, ensuring that the system remains responsive to new research findings, policy developments, and evolving sustainability trends.



\begin{acks}
This project is supported by Swiss National Science Foundation (SNSF) under grant no. \href{https://data.snf.ch/grants/grant/10000100}{10000100}. We thank Rahel Meili for her constructive feedback for SwissSPC. 
\end{acks}


\bibliographystyle{ACM-Reference-Format}
\bibliography{custom}









\end{document}